\documentclass[runningheads]{llncs}
\usepackage{times}
\usepackage{epsfig}
\usepackage{graphicx}
\usepackage{amsmath}
\usepackage{amssymb}
\usepackage[utf8]{inputenc}
\pdfoutput=1
\usepackage{dblfloatfix}
\usepackage{float}

\usepackage{fourier}

\usepackage[pagebackref=true,breaklinks=true,colorlinks,bookmarks=false]{hyperref}

\begin{document}

\title{Lowering Detection in Sport Climbing Based on Orientation of the Sensor Enhanced Quickdraw}

\author{Sadaf Moaveninejad \inst{1}\orcidID{0000-0003-2144-4412} \and
Andrea Janes\inst{2}\orcidID{0000-0002-1423-6773} \and
Camillo Porcaro\inst{1}\orcidID{0000-0003-4847-163X}
}

\authorrunning{S. Moaveninejad et al.}
\institute{Department of Neuroscience and Padova Neuroscience Center, University of Padova, Padova, Italy\\
\email{\{sadaf.moaveninejad, camillo.porcaro\}@unipd.it}
\and
FHV Vorarlberg University of Applied Sciences, Dornbirn, Austria\\
\email{andrea.janes@fhv.at}
}
\maketitle

\begin{abstract}
   Tracking climbers' activity to improve services and make the best use of their infrastructure is a concern for climbing gyms. Each climbing session must be analyzed from beginning till lowering of the climber. Therefore, spotting the climbers descending is crucial since it indicates when the ascent has come to an end. This problem must be addressed while preserving privacy and convenience of the climbers and the costs of the gyms. To this aim, a hardware prototype is developed to collect data using accelerometer sensors attached to a piece of climbing equipment mounted on the wall, called quickdraw, that connects the climbing rope to the bolt anchors. The corresponding sensors are configured to be energy-efficient, hence become practical in terms of expenses and time consumption for replacement when using in large quantity in a climbing gym. This paper describes hardware specifications, studies data measured by the sensors in ultra-low power mode, detect sensors' orientation patterns during lowering different routes, and develop an supervised approach to identify lowering. 
\end{abstract}

\vspace{-0.2cm}
\textbf{Keywords}
Machine Learning, Artificial Intelligence, Internet of Things.
\vspace{-0.1cm}
\section{Introduction}
Sport climbing has been recently receiving increased popularity internationally not only as a recreational sport but also as a competitive one ~\cite{ivanova2020video}. This sport could be carried out either outdoor on natural cliffs or indoor in climbing gyms. In case of indoor, climbing gyms provide walls consist of a number of lines, while each line in turn comprises several routes with various difficulties. In addition, climbing gyms must provide the climbers with equipment and services which gives them the  opportunity to challenge themselves and improve their skills without scarifying their safety. To this aim and similar to other sports (\textit{i.e.} cycling, running), engineering and science have begun to help sport climbing ~\cite{international2006engineering}. 

The contribution of science in sport activities could takes place in the form of collecting data from athletes using sensors embedded in electronic equipment such as smart watches, smart bands, fitness trackers, and smart phones. These devices are attached to the body of the athletes (as in \cite{Boulanger2016,Ladha2013,Kosmalla2015,Pansiot2008,boulanger2015automatic} ). The corresponding measurements are time-series data used to visualize statistics of athletes and analyze their performance. Such analysis could be done either by a coach or trough certain applications. Another way of acquiring data to analyze sport activities, is based on camera. In this approach, computer vision algorithms are developed to extract human 2-dimensional (2D) pose sequences from video frames and based on skeleton estimation of each person \cite{einfalt2019frame,sasaki2020exemposer}. Instrumented climbing walls is the other approach for monitoring climbers which is inspected in \cite{aladdin2012static,fuss2008instrumented}

The data acquisition methods based on wearable-sensors and camera are not always desirable for climbers. The former limits convenience of the user when wearing an extra device during climbing, and the latter is against the privacy. Keeping these issues in mind, and similar to \cite{ivanova2020video,moaveninejad2024climbing}, for this paper data is collected from accelerometer sensors attached to a piece of climbing equipment mounted on the wall, called quickdraw. The sensor enhanced quickdraw, hereinafter is referred to as smart-quickdraw (s-qd). To the best knowledge of the authors, the sensor enhanced quickdraws were studied the first time in \cite{ivanova2020video}. Later in \cite{moaveninejad2024climbing}, we utilized sensors in ultra-low power mode attached to the quickdraws, to detect patterns in data during climbing different routes. Using sensors in ultra-low power mode was due to the fact that apart from climbers, another concern for data collection relates to the climbing gyms. There is a large number of quickdraws in a climbing gym and regular changing batteries of the s-qds is expensive for the gym in terms of both time and costs. Hence, sensors attached to the quickdraws must be energy efficient such that their batteries do not need to be replaced in the short term. 

The data collected from climbers helps both climbers and gyms in different ways. One goal is to improve performance of the climbers, to achieve this, first must detect different activities during climbing \textit{i.e.} ascending, resting, falling, lowering, and rope-pulling at the end of the climb. Then, the detected activities could be analyzed by expert and in off-line modes to provide recommendations for a better performance, or in on-line mode to improve safety by early prediction of a risk. Another goal is to improve infrastructure of the gyms based on climbers' needs. 

In this work, we investigated on data from the lowest s-qds on the wall to detect the lowering. Lowering refers from the time climber finishes clipping the carabiner and starts to come down, till he reaches to the ground and leave the rope then the belayer can start pulling and collecting the rope \cite{ivanova2020video}. This reveals the end of each climbing activity and helps to separate sequential climb to be analyzed one by one. Authors in \cite{ivanova2020video} introduced a system to hybrid system based on sensor and camera was developed to detect rope-pulling activity. Our proposed method is a supervised algorithm based on classification of samples measure from a single sensor at the lowest position of the line.

However, using sensor in energy-efficient mode and attaching them to the wall, instead of the body of the climber, introduces some challenges. Saving energy reduces number of sensor sample and introduces challenges in training machine learning models. Moreover, different from algorithms based on wearable devices which mainly collect data through a single sensor on the body of the climber, we need to deal with sensors when they are on the wall. In the following, all these points are addressed in details.

\subsection{Contributions}
One 27 years old male climber participated in the experiment. He has been climbing for thirty years with advanced skill level i.e., self-estimated as a and 6b on-sight on the French Rating Scale of Difficulty (FRSD). For the purpose of data collection, the participant was asked to climb three routes in the leading style in two different days. The three pre-selected routes were with difficulty levels of 5c+,6a+,and 6b+ (in French grading system) and selected according to the skill level of the climber.  The participants climbed in his usual speed, clipping the rope into every quickdraw, and was free to take resting time between climbs. The climber pulled the rope after lowering and there was another person responsible for belaying. 

In the first day, he climbed each route five times. The order of climbing was (5c+, 6a+, 6b+) and repeated this process five times. After three weeks, we asked climber to repeat a similar experience but the order of climbing was: (5c+, 5c+, 5c+), (6b+, 6b+, 6b+), and
(6a+, 6a+, 6a+). The aim of changing the order of routes was to add different parameters \textit{i.e} tiredness, and provisioning to the data collected from the same routes. 

All eight quickdraws in the line were enhanced with sensors. Climber started climbing by attaching the rope to the first quickdraw, the ascending ends after clipping the rope to the quickdraw number eight and lowering started.

Employing sensors in ultra-low-power mode, enhances energy saving, but instead significantly reduces the number of samples. Considering this fact, we need to carefully understand the sensor functionality to analyze the received samples and find the more informative ones for our objective. To this aim, we described the details of our data acquisition system in Sec.~\ref{Sec:Data Acquisition System}. Afterwards in Sec.~\ref{Sec:Features engineering}, the relevant feature for lowering detection are extracted from sensor measurements.  In Sec.~\ref{Sec:Classification} the selected features are used for detect lowering and the corresponding results are discussed.
\section{Data Acquisition System}
\label{Sec:Data Acquisition System}
\begin{figure}[h]
    \centering
    \vspace{-0.3cm}
    \includegraphics[width=0.7\textwidth]{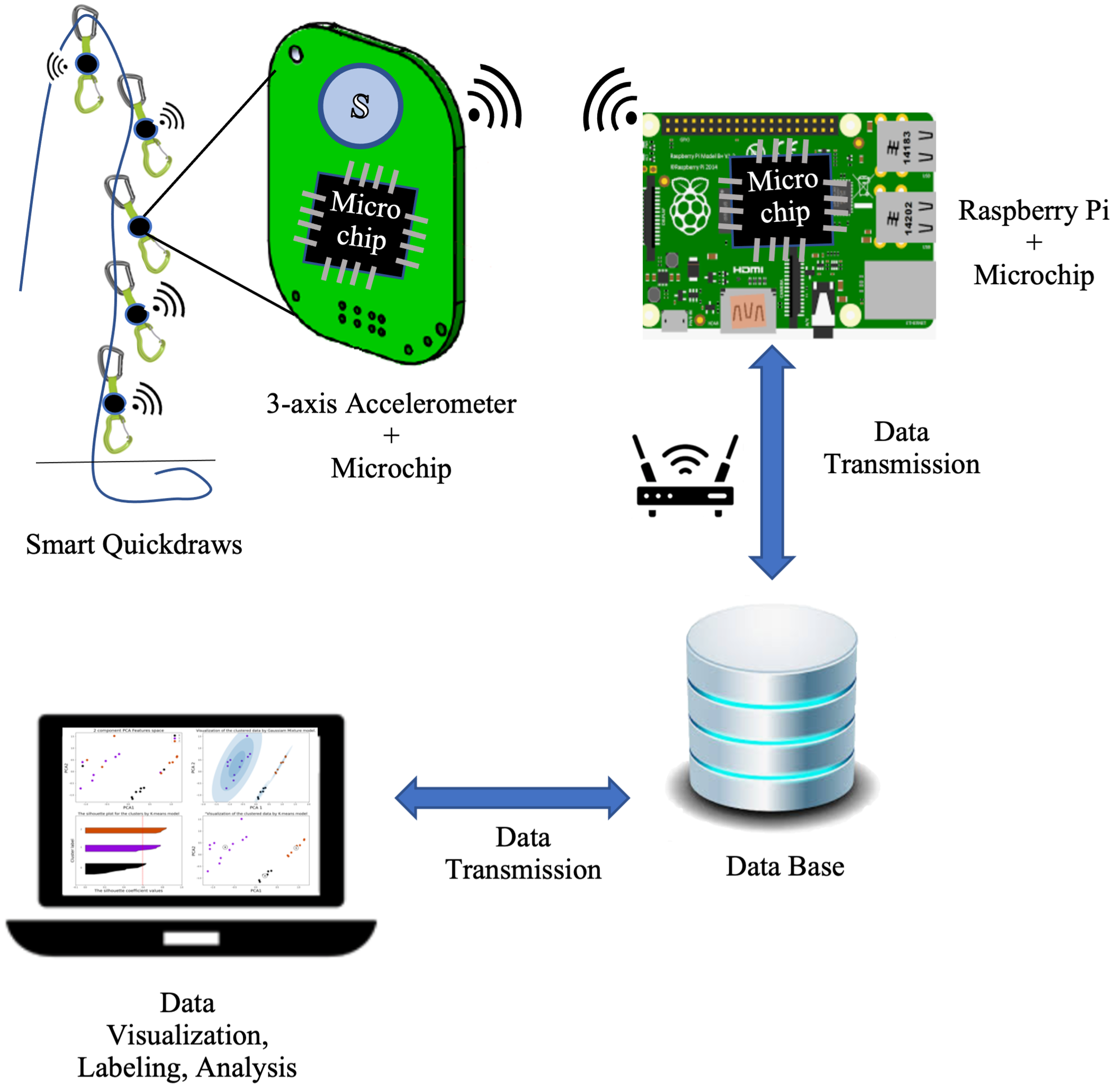}
    \caption{Data acquisition system}
    \vspace{-0.5cm}
    \label{fig:iot}
\end{figure}

Our data acquisition system in shown in Fig \ref{fig:iot} and consists of two main parts: 1- smart-quickdraws to measure 3-axis accelerations, 2- a base station to receive data from sensors and forwards them to the database in a remote server.
\subsection{Smart-quickdraw}
The so called smart-quickdraw essentially refers to an in-house circuit board which is attached to the strip in the central part of the quickdraw. This board consists of an accelerometer sensor to capture quickdraw movements and a microchip to control the accelerometer and communicate with the base station. For our application, energy efficiency is a primary issue as battery replacement of all smart-quickdraws in a climbing gyme is both time and cost consuming. In this regard, for the accelerometer we selected LIS3DH by STMicroelectronics which could be configured to operate in ultra-low-power modes through smart sleep-to-wake-up and return-to-sleep functions. Concerning the microchip, we used ATSAMR21G18A by Atmel which combines an microcontroller unit (MCU) and a RF transceiver. The accelerometer sensor is controlled and configured via a firmware which is developed inside the MCU, moreover the smart-quickdraw communicates with the base station via transceiver. In the following, some of our main settings for the accelerometer sensors are listed: 
\begin{itemize}
    \item {\bf Full scale:} 
    The LIS3DH has dynamic user selectable range of forces it can measure which are from ±2g to ±16g. Typically in accelerometers, the smaller the range, the more sensitive the readings will be. Hence, we selected 2g as the full scale for our prototype.
    \item {\bf Data rate:}
    The LIS3DH is capable of measuring accelerations with output data rates from 1 Hz to 5 kHz. For our energy efficient system, we selected 10 Hz and 50 Hz for the inactive and active modes, respectively.
    \item {\bf Output bits:} 
    The analog to digital converter (ADC) of LIS3DH could have 8, 10, or 12 bits data output for operating in low-power, normal and high-resolution modes. Since our goal is to have energy efficient system, we selected 8-bit resolution for the sensor data output.
    \item {\bf Range and resolution:}
    Accelerometers usually provide raw values which are not equivalent to meters per second squared. Consequently, we still need to scale the accelerometers' out put based on our setting for full scale. In this sensor's case, a signed 8-bit output data corresponds to a number ranging from -127 and 127. After scaling this range by ±2g full scale, sensor data output 127 is +2g of force, and -127 is -2g. The resolution of the data output after scaling is $\frac{2g}{127} \approx 16 mg$. Hence, knowing the range and scale is a key to deciphering the sensor's data output.
    \item {\bf Noise reduction:}
    At a sampling rate of 50 Hz there is still considerable noise in the sensor readings, which is the reason why a software averaging method has been implemented in our prototype. This simple method takes every subsequent set of 8 samples and calculates the average for the acceleration on each axis. Then, the corresponding averaged value is transmitted to the raspberry-pi.
    \item {\bf Filter insignificant changes:}
    Regardless of whether there is movement or not, the accelerometer always provides sensor readings for the configured sampling rate. Since bandwidth is a premium for low power devices, sensor values without any changes on any axis are not transmitted to the base station. However, even after averaging, hardly any sample is the same except if the sensor is completely still. Therefore a method has been implemented that prevents samples to be sent if there is no absolute change on any axis below a certain threshold. For this prototype, we set a threshold value equal to 15 units of sensor's data output. The sensor does not consider any sample worth transmitting if the measure value of none of the axis has changed by at least 15 units. As mentioned above, the resolution of the data output for this sensor configuration is almost 16mg, therefore the 15 units threshold is equivalent to the acceleration of $15 \times 16\ mg = 240 \ mg$.
    \item {\bf Grouping samples:}
    To save even more power (the radio of the smart-quickdraw is always off while no data is being sent) all sensor values are sent to the base station as a batch of values, i.e., grouped. We set this value to two samples, meaning that no sample is sent out individually, except prior to going to sleep after inactivity, in that case any sample that was held back is being sent out. This may have the effect that sensor data is not received to the base station in real-time, but with a delay. However, since every sensor value has a timestamp this should not be an issue for analysis use-cases.
\end{itemize}
\subsection{Base station}
The base station of our system comprises a board with the microchip, similar as in the smart-quickdraw, mounted on a raspberry-pi. The raspberry-pi is used as a powerful cpu and hosts the application programming interface (API) that allows to communicate with the devices, as well as a web-server and possible other services. In addition, the raspberry-pi cpu communicates with its board via the serial line. The firmware of the board within the base station performs the "network translation" to the radio network that is used amongst the boards (inside the base station and smart-quickdraws).

\subsection{Ultra-low-power data acquisition:}
\label{subsec:Ultra-low-power data acquisition}
To enhance power saving the sensor is configured, through the cpu of the smart-quickdraw (within MCU), to switch between sleep and active modes as below. Note that cpu of the base station does not interact with the sensor and perform any configuration steps on it. 
\begin{itemize}
    \item cpu is in sleep mode, sensor is not moving and it samples data with the sampling rate of sleep mode which is 10 Hz, 
    \item sensor data output is compared with a threshold value configured in the firmware of the smart-quickdraw.
    Then, one of the following situations happens:
    \begin{enumerate}
        \item the measured values in all the 3 axis are lower than the threshold (15 units of sensor's data output). In this case, the samples are filtered and not read by the cpu 
        \item the sensor data of at least one of the 3 axis exceeds the activation threshold. If so, cpu re-configures the sensor to the sampling rate of run mode which is 50 Hz and starts reading its values 
    \end{enumerate}
    \item when sensor is in active mode, the average of each 8 samples is calculated and transmitted to the base station. Smart-quickdraw keeps on comparing its data with the threshold, in case the averaged sample goes below threshold for more than 0.8 seconds sensor is considered as inactive. If sensor is inactive for a duration of 20 second, cpu reconfigures it to the sampling rate of sleep mode and stops reading its data 
\end{itemize}

\section{Features Engineering}
\label{Sec:Features engineering}
The sensor embedded in smart-quickdraw works in ultra-low-power mode, hence the s-qd does not continuously send data to the base station. Instead, transmits samples when it is in active mode and the change in the movement of the corresponding quickdraw exceeds a certain threshold. To detect lowering, in our experience the first two sensors of the line are sensor enhanced and $i$ refers to the position of the s-qds in the line. During climb $c$, the sensor at position $i$ transmits samples with timestamp $t_{{i}_{k}}$:
\begin{equation}
{T_{i}}^{c} = \left \{ t_{{i}_{k}} | k\geqslant 1 \right \}
\end{equation}
where $k$ refers to the index of the samples and ${T_{i}}^{c}$ is the set of timestamps of all samples transmitted from s-qd at position $i$ to the base station. These samples are measured from the time when the s-qd is clipped till the climber's movement still causes tangible changes in the acceleration of the s-qd. 

After visually keeping track of the climbers in gym, we assume that the most relevant information for detecting lowering is the orientation of the first s-qd which the climber clips to the rope. This quickdraw could be either at the first or second position in the line. Accordingly, first we calculate orientation of the first clipped s-qd in three planes and then extract statistical features from orientation in each of the three planes \textit{i.e.} mean, max, standard deviation etc. 

\subsection{Orientation of s-qd}
\begin{figure}[t!]
    \centering
    \vspace{-0.3cm}
    \includegraphics[width=0.35\textwidth]{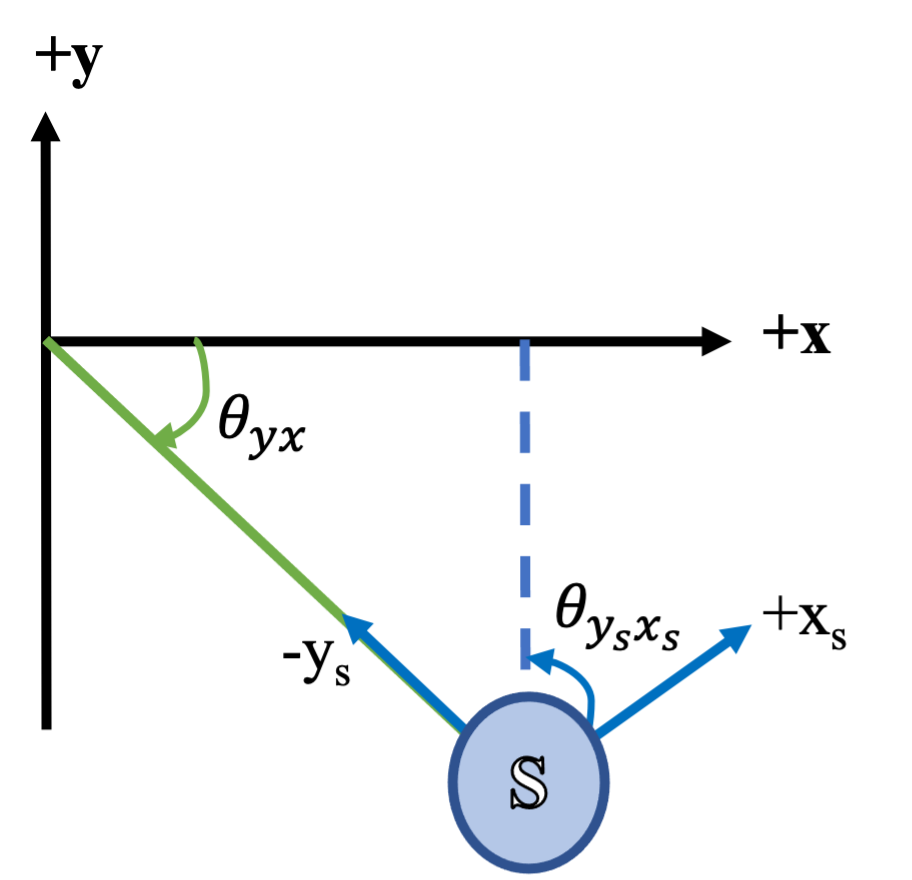}
    \caption{x,y, z directions of 3-axis accelerometer sensor with respect to the climbing wall}
    \vspace{-0.5cm}
    \label{fig:teta}
\end{figure}
\begin{figure*}[b!]
    \centering
    \vspace{-0.5cm}
    \includegraphics[width=0.9\textwidth]{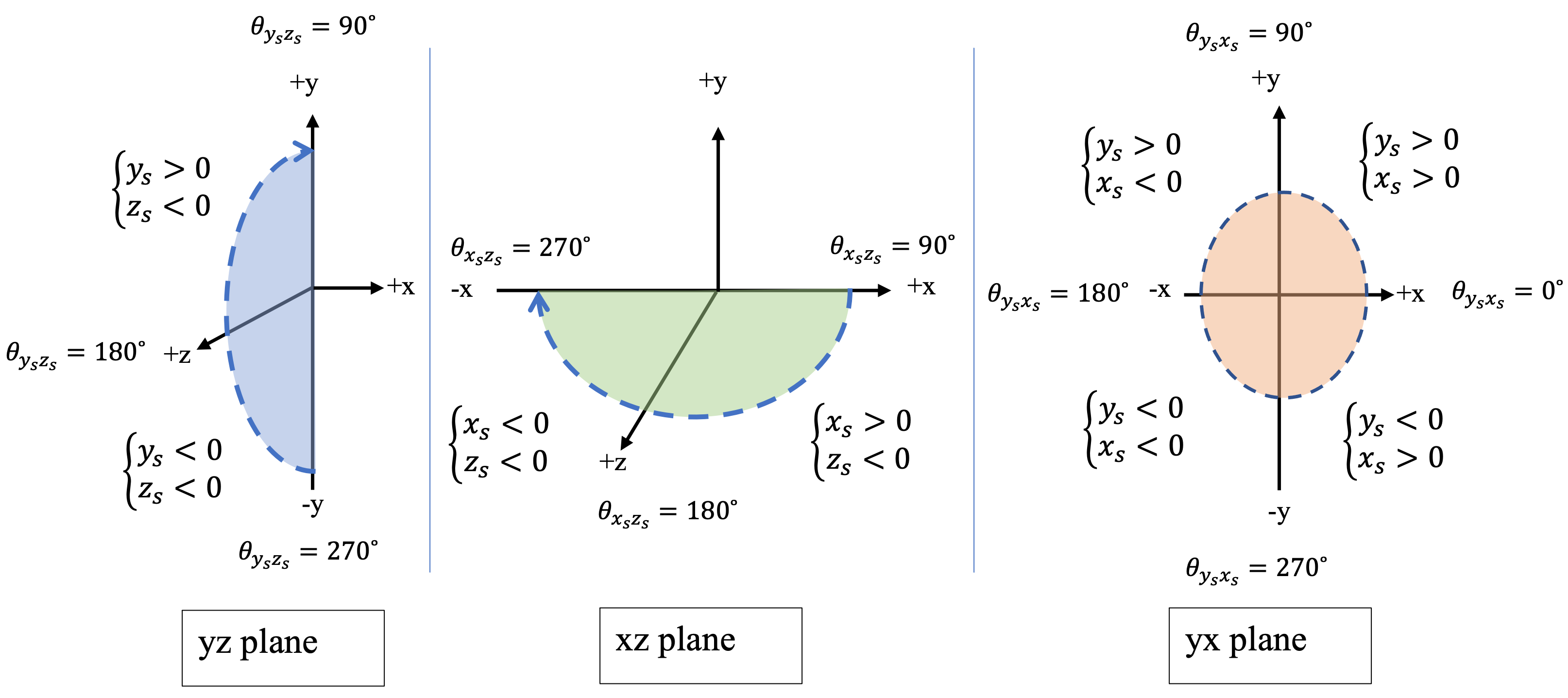}
    \vspace{-0.2cm}
    \caption{Mapping orientation of a smart-quickdraw to the wall}
    \vspace{-0.5cm}
    \label{fig:2d_planes}
\end{figure*}
The feature that which can provide us meaningful information about position of the s-qd and helps us to distinguish the lowering activity is the orientation of the quickdraws during climbing. To use such feature, we need to calculate orientation of the quickdraw from 3-axis acceleration measured by its sensor. To this aim, we evaluate the discrete orientation of the s-qd for each sample independent of the prior and posterior ones. Moreover, samples are analyzed in three 2-dimensional (2-D) coordinate plans to detect whether the s-qd is upward or downward, right or left, forward or backward. Hence, for each sample we have $3$ values $(\theta_{y_{s}x_{s}}, \theta_{y_{s}z_{s}}, \theta_{x_{s}z_{s}})$ correspond to the orientation of the sensor in ${y_{s}x_{s}}$-plan, ${y_{s}z_{s}}$-plan, and ${x_{s}z_{s}}$-plan, respectively. Here, $(x_{s}, y_{s}, z_{s})$ refers to the 3-axis acceleration of the sensor in its own coordinate system. Orientations of s-qd with respect to the coordinate system of the sensor must be mapped to the coordinate of the wall $(x, y, z)$, as shown in Fig \ref{fig:teta}. thus:
    \vspace{-0.2cm}
    \begin{equation}
    \theta_{yx} = \theta_{y_{s}x_{s}}, \ arctan(\frac{-y}{x}) = arctan(\frac{-y_{s}}{x_{s}})
    \end{equation}
    \vspace{-0.2cm}
    \begin{equation}
    \theta_{yz} = 180^{\circ} - \theta_{y_{s}z_{s}}, \ arctan(\frac{y}{z}) = arctan(\frac{y_{s}}{-z_{s}})
    \end{equation}
    \vspace{-0.2cm}
    \begin{equation}
    \theta_{xz} = 180^{\circ} - \theta_{x_{s}z_{s}}, \ arctan(\frac{x}{z}) = arctan(\frac{x_{s}}{-z_{s}})
    \end{equation}
The outcome of such conversion between coordinates systems is given in Fig \ref{fig:2d_planes}. For each s-qd, $\theta_{y_{s}x_{s}}$ and $\theta_{x_{s}z_{s}}$ are enough to reveal whether the s-qd is oriented 
    up-ward, down-ward, right, left, or middle. 

\subsection{Orientation of s-qd during lowering}
\begin{figure*}[t!]
    \centering
    \vspace{-0.2cm}
    \includegraphics[width=\textwidth]{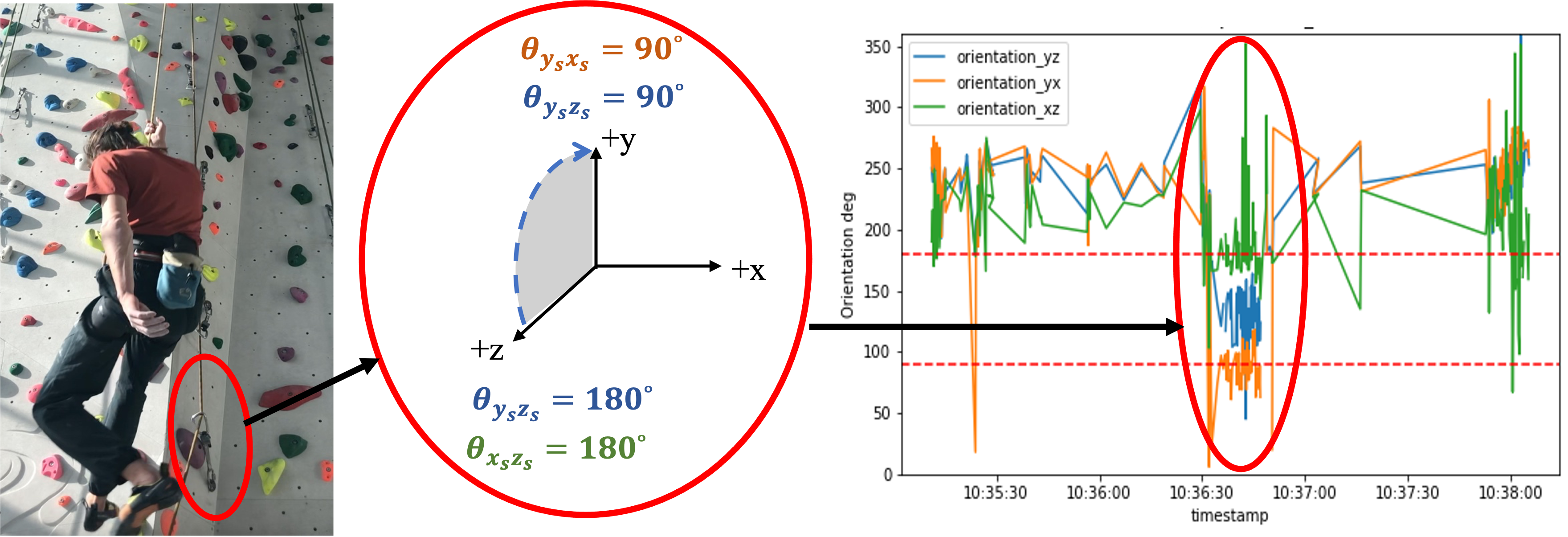} 
    \caption{Orientation of the first smart-quickdraw during lowering}
    \vspace{-0.2cm}
    \label{fig:lowering}
    \end{figure*}
We can use orientation to detect lowering and consequently the end of a climb. Lowering is an activity which lasts several seconds, in one side there is belayer on the ground holding the rope, on the other side the climber is hanging on rope while being lowered. Hence, tension applied to the rope from climber is more than the belayer. Consequently, as long as the climber is lowering, the first quickdraw has upward orientation orthogonal to the wall as seen in Fig \ref{fig:lowering}. In other words, the orientation of the first s-qd must be between ($90^{\circ}$- $180^{\circ}$) in the ${y_{s}z_{s}}$-plan, around $90^{\circ}$ in ${y_{s}x_{s}}$-plan, and around $180^{\circ}$ in ${x_{s}z_{s}}$-plan. keeping this in mind, the lowering is evident from sensor measurements in \ref{fig:lowering}. It worth mention that due to the presence of the wall, the orientation of the s-qd in ${x_{s}z_{s}}$-plan is limited between ($90^{\circ}$- $270^{\circ}$) and the values out of this range is a sign of change in coordinates of the sensor which is caused by twisting. 

On this assumptions, a vector of orientation could be calculated from samples measured by the lowest s-qd in a period between the moment when s-qd is activated through clipping till it is still and stops sampling. 
\subsection{Resampling}
\begin{figure}[b!]
    \centering
    \vspace{-0.2cm}
    \includegraphics[width=0.55\textwidth]{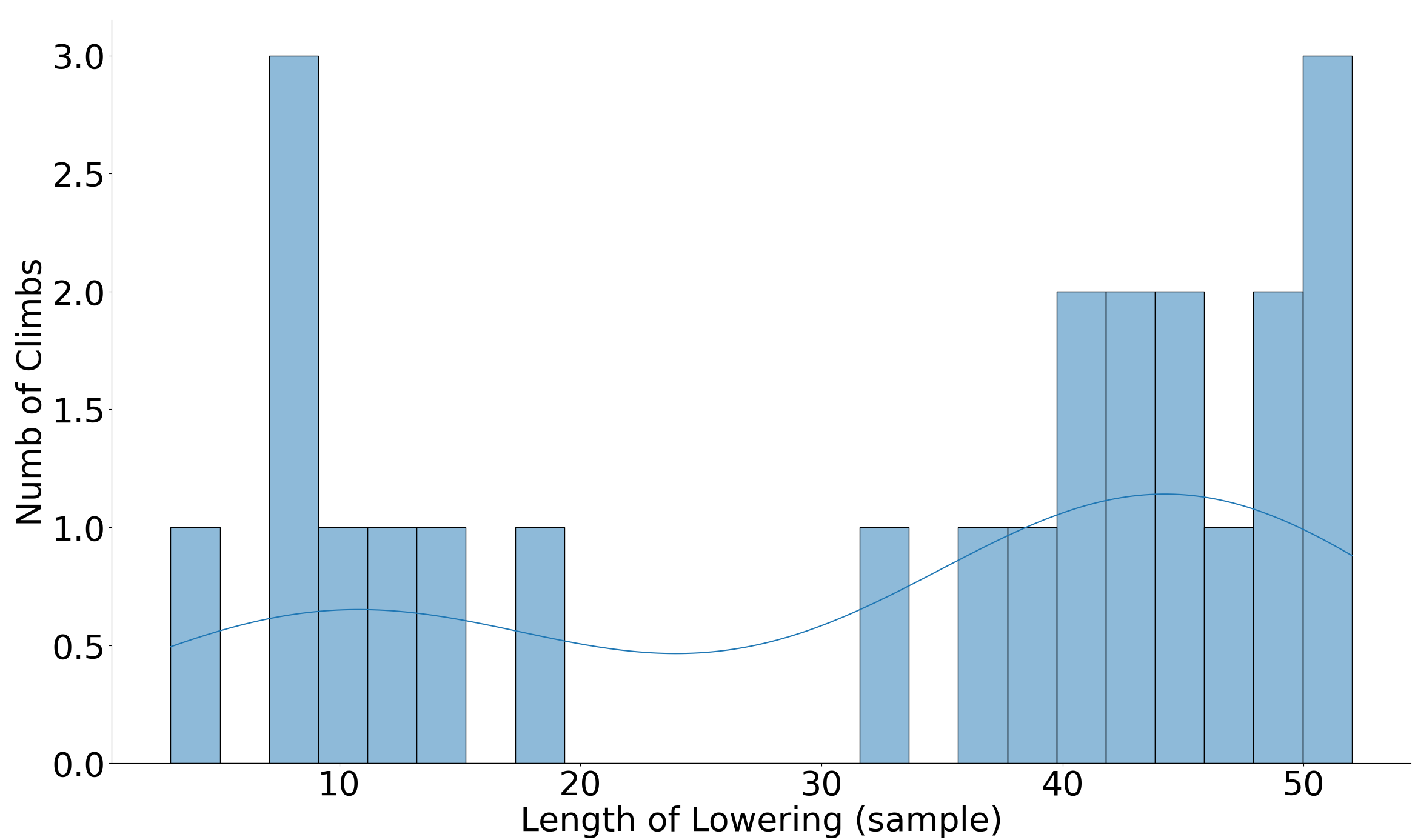}
    \caption{Duration of lowering in different climbs before resampling}
    \vspace{-0.5cm}
    \label{fig:lowering_hist}
    \end{figure}

\begin{figure}[t!]
    \centering
    \vspace{-0.2cm}
    \includegraphics[width=0.55\textwidth]{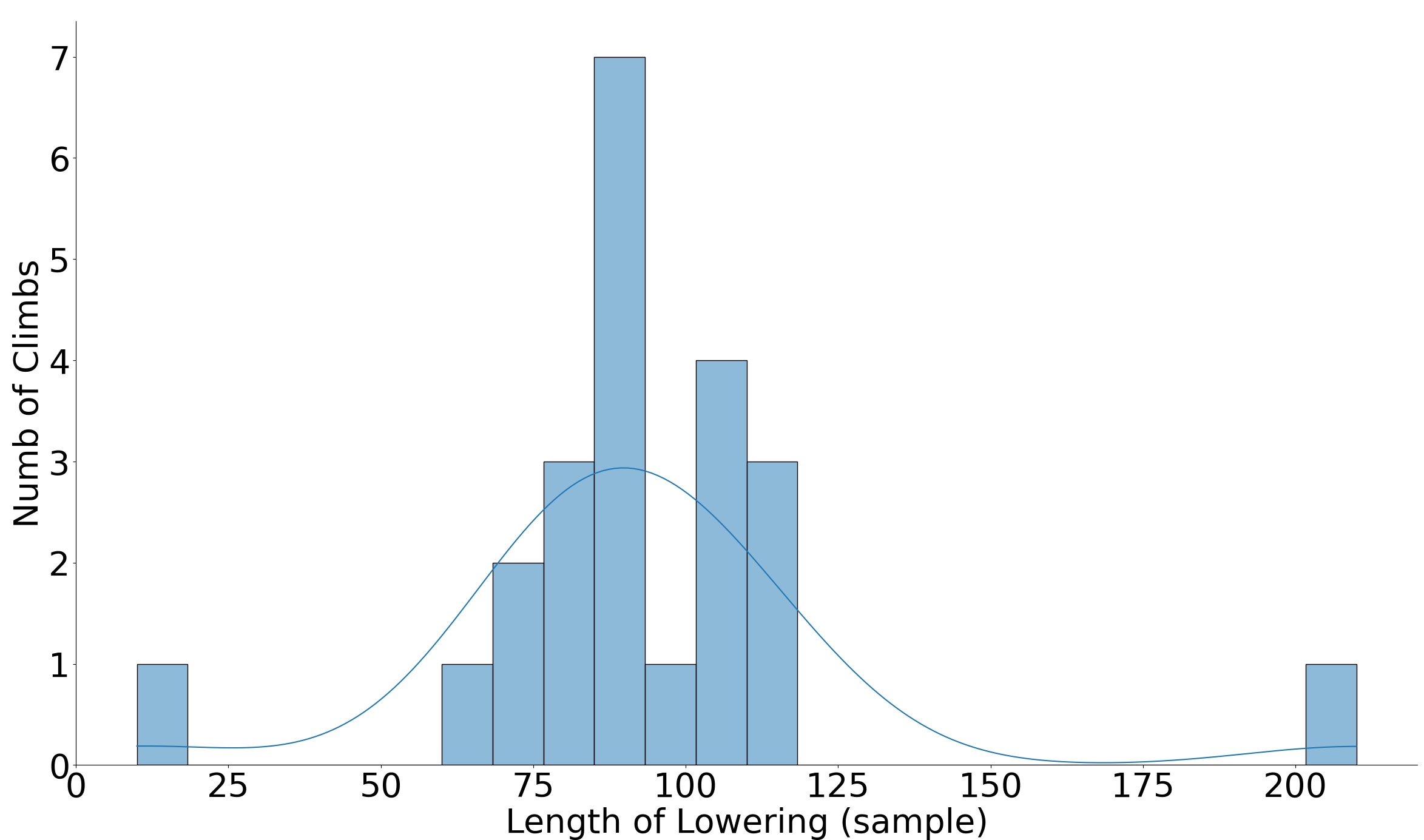}
    \caption{Duration of lowering in different climbs after resampling}
    \vspace{-0.5cm}
    \label{fig:lowering_hist_resampled}
    \end{figure}

Different climbs have different duration and different number of samples. Therefore before classification we use resampling to have the same number of samples for all climbs. Figures \ref{fig:lowering_hist} and \ref{fig:lowering_hist_resampled} shows number of samples during lowering before and after re-sampling the climbs to one minuet respectively. Before classification, features are optimized in two steps: 1- feature scaling, 2- feature selection. 
\section{Classification:}
\label{Sec:Classification}
\begin{figure}[!b]
    \centering
    \vspace{-0.4cm}
    \includegraphics[width=0.9\textwidth]{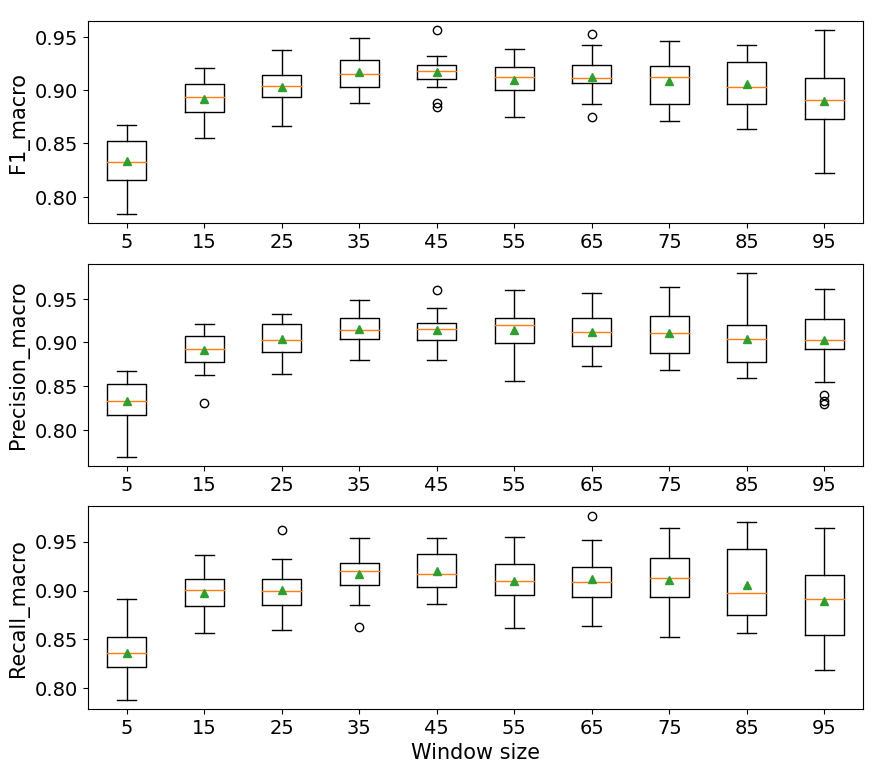} 
    \vspace{-0.2cm}
    \caption{Classification of lowering and not-lowering for different window sizes}
    \vspace{-0.5cm}
    \label{fig:classification_lowering}
\end{figure}
lowering activity is a temporal event that may be detectable from data measured through sensors. In this regard, sliding windows are used to transform each interval of time-series data into an appropriate feature vector (orientation along 3-axis and corresponding statistics). As a result, samples from s-qd during each climb is converted to several sliding windows with or without overlap.

For classification problems, one typically uses stratified k-fold cross-validation, in which the folds are selected so that each fold contains roughly the same proportions of class labels. In repeated cross-validation, the cross-validation procedure is repeated n times, yielding n random partitions of the original sample. The n results are again averaged (or otherwise combined) to produce a single estimation.

In this work, we used stratified 10-fold cross validation with 3 times repetitions over sliding windows with different lengths to find the optimum length for windows which corresponds to the best performance of the classifier. Regarding the classifier, a decision tree is used. 

We assigned the label for a windows to be \textit{lowering} only if the $90\%$ of its samples are labeled as lowering otherwise the window is labeled as \textit{not\_lowering}. The subsequent windows have 2 samples overlap. Figure \ref{fig:classification_lowering} shows that the best results (precision, recall, and F1 scores more than $90\%$) corresponds to window length equals to 45 samples and afterwards the performance of the classifier doesn't improve. As shown in Fig \ref{fig:lowering_hist_resampled}, after resampling, the length of lowering in almost all climbs are more than 60 samples. the optimum window length (45 samples) is equal to the half of the number of samples during lowering after resampling.
\section{Conclusion}
\label{Sec:Conclusion}
In this work, we utilized 3-axis accelerometer sensors working in ultra-low power mode and attached them to the quickdraws hanging from climbing walls. Features are extracted from sensor at the lowest position measured while a climber was climbing 24 times the same line consists of 3 routes, in two different days.

From each measured sample, orientation is calculated in 3-D plane. The statistical features obtained from orientation of s-qd used for classification shows that lowering activity is detected with precision more than $90\%$ using orientations of s-qd. This results are aligned with what we observed visually in climbing gyms, when the first s-qd was oriented upward all the time during lowering. This is due to the difference between tension of the rope from up (climber) and down (belayer).
    
    

\section*{Acknowledgments}
This work has been partly supported by the project ``Sensors and data for the analysis of sport activities (SALSA)'', funded by the EFRE-FESR programme 2014-2020 (CUP: I56C19000110009). 
\bibliographystyle{splncs04}
\bibliography{references}
\end{document}